# Morphology of Galaxy Mergers at Intermediate Redshift

J. Christopher Mihos

*Board of Studies in Astronomy and Astrophysics,
University of California, Santa Cruz, CA 95064*

hos@lick.ucsc.edu

## ABSTRACT

Using a combination of numerical simulation and synthesized Hubble Space Telescope Wide Field Camera 2 (HST WFC2) images, we follow the detectability of morphological signatures of disk galaxy mergers at intermediate redshifts $z = 0.4$ and $z = 1.0$. Rapid evolution in the surface brightness of tidal tails makes their use as an interaction signature limited; long tidal tails will only be detectable relatively early in the encounter, before the galaxies actually merge. Violent relaxation rapidly smooths isophotal irregularities after the galaxies merge, leaving the very low surface brightness tidal debris surrounding the remnant as the only sign of peculiarity after a few hundred Myr. This debris becomes indistinguishable from the main body of the remnant after 0.5 (z=1.0) to 1 (z=0.4) Gyr, leaving a morphologically normal elliptical galaxy behind. Using WFC1, *all* signatures become undetectable almost immediately once the galaxies merge, due to the aberrated PSF. We interpret these results in light of recent HST programs investigating the nature of starburst and E+A galaxies in moderate redshift clusters.

*Subject headings:* galaxies:interactions, galaxies:evolution, galaxies:structure, galaxies:peculiar, galaxies:general

## 1. Introduction

Mergers and interactions between galaxies are thought to drive substantial galaxy evolution, triggering starburst and AGN activity and transforming spiral galaxies into ellipticals and S0's. Anomalous colors and stellar populations in galaxies are often interpreted as being the result of merger- or interaction-induced starbursts (e.g., Larson & Tinsley 1978), especially in dense clusters of galaxies. In fact, the morphology-density relation itself has been ascribed to the effects of mergers in high density environments (see Oemler 1992 for a review). However, observational data regarding the direct effects of mergers, crucial for testing such evolutionary scenarios, has been limited primarily to the local universe, at $z \lesssim 0.1$, providing a very limited baseline over which to examine galaxy evolution.

To extend this baseline of observations to higher redshift, recent studies have used deep Hubble Space Telescope imaging of galaxies at intermediate redshift ($0.1 \lesssim z \lesssim 1.0$) to examine the morphological type of both field (Griffiths et al. 1994a,b) and cluster (Dressler et al. 1994a,b) galaxies. The high spatial resolution and reduced sky background offered by HST allows for accurate morphological typing of such galaxies, often revealing features such as multiple nuclei, tidal features, and distorted isophotes, possibly indicative of a merger event. Using such morphological features, the effects of galaxy mergers at intermediate redshift may *in principle* be probed. For example, Forbes et al. (1994) find that interacting/merging field galaxies have somewhat bluer nuclear colors than non-interacting systems; in rich clusters, such interactions may be responsible at least in part for the Butcher-Oemler effect. For the anomalous "E+A" systems (galaxies with spectral evidence for both an old stellar population and a recent starburst), the situation is less clear. In a recent investigation of the intermediate redshift clusters AC114 ($z = 0.31$) and Abell 370 ($z = 0.37$), Couch et al. (1994) found little or no evidence for merger signatures among E+A galaxies, suggesting other mechanisms may be responsible for triggering bursts of star formation in otherwise normal ellipticals. However, the E+A galaxies in the more distant cluster 0016+16 ($z = 0.55$) seem to be poorly described as pure ellipticals (Wirth, Koo, & Kron 1994), implying that the formation of E+A systems may be a varied process.

A major complication to such approaches is that the morphological signatures of a merger will disappear much more quickly than the spectrophotometric signatures of any starburst population born in the merging process. As shown in the calculations by Barnes (1988, 1992) and Hernquist (1993a,b), tidal tails and bridges formed during the initial collision rapidly disperse, making them very difficult to detect at late times. The situation is no better in the inner regions; although the much higher surface brightness makes detection of peculiarities easier, the violent relaxation and phase mixing associated with the merger rapidly homogenize the remnant, smoothing out any isophotal irregularities on short timescales. Hence the morphologically smooth E+A galaxies in AC114 and Abell 370 (Couch et al. 1994) may in fact be merger remnants of intermediate age. In order to understand the selection effects and incompleteness associated with morphological classification at these redshifts, therefore, the timescale over which such features are detectable must be quantified.

In this paper we address the question of detectability of merger signatures at these intermediate redshifts. Starting with numerical models of merging disk/bulge/halo galaxies, we model the appearance of the merging galaxies at various phases of their evolution at redshifts of $z = 0.4$ and $z = 1$ using the characteristics of the HST WFC2 CCDs. We perform both a qualitative image classification and a quantitative isophotal analysis to examine the longevity of morphological signatures of mergers as they would appear in HST imaging. We discuss these images in the context of recent HST observations of moderate redshift clusters.

## 2. Numerical Technique

Complete details of the merging galaxy models used in this analysis can be found in Hernquist (1993), and we give only a brief description here. The model galaxies consist of spherical, isothermal dark matter halos, exponential stellar disks, and oblate central bulges. The system of units employed defines $G = 1$, the scale length of the exponential disk $h = 1$, and the disk mass $M_d = 1$. Scaling to values typical of the Milky Way, unit length is 3.5 kpc, unit mass is $5.6 \times 10^{10}$ $M_\odot$, and unit time is $1.3 \times 10^7$ yr. Halos have total mass $M_h = 5.8$ and core radius $\gamma_h = 1.0$, while the bulges have mass $M_b = 1/3$, major axis scale length $a = 0.2$ and minor axis scale length $b = 0.1$. In the models analyzed here, the disks and halos consist of 65,536 particles each, while the bulges consist of 16,384 particles each.

Given the enormous range of possible interaction parameters, rather than attempting an exhaustive survey of mergers, we choose to focus the analysis on a fiducial example. In this example, which corresponds to Hernquist's (1993) model 4, two identical disk galaxies are placed on a parabolic orbit such that the closest approach in the Keplerian orbit would be $R_{peri} = 2.5$ at a time $t = 24$. The orientations of the disks with respect to the orbital plane are given by their inclination $i$ and argument of pericenter $\omega$. Both galaxies are highly inclined, with $(i, \omega) = (71°, 30°)$. Shortly after the initial passage, the disk galaxies form prominent tidal bridges and tails which expand quickly away from their parent galaxies. The galaxies continue to separate, but the efficient transfer of orbital energy to internal energy of the galaxies results in a braking effect, and the galaxies turn around and merge on a timescale of $\sim 500$ Myr after the initial encounter.

The synthesized HST images are constructed by first binning the three dimensional model onto a two dimensional surface, a task which involves a choice of both viewing geometry and pixel size. To avoid confusion between the projected separation of the galaxies and their true separation, we choose the viewing angle such that the system is observed projected onto the orbital plane. To then assign the pixel size for binning the model, choices of cosmology must be made. As an illustrative example, we chose an $H_0 = 50$ km/sec/Mpc, $q_0 = 1/2$ universe. In such a universe, at a redshift of $z = 0.4$ the 0″.1 WFC2 pixel corresponds to a spatial scale of $\sim 1.25$ kpc; at $z = 1.0$, 0″.1 corresponds to 3.4 kpc.

To convert the mass per pixel of the synthesized images into luminosity per pixel, a mass-to-light ratio must be adopted for the particles comprising the luminous portions of the galaxies. This mass-to-light ratio is very different from the mass-to-light ratio for the disk galaxies as a whole, which would include contributions from the dark matter halos (if any). Kent (1985) finds that for the disk component of spiral galaxies, M/L $\sim 2.5$ M$_\odot$/L$_\odot$ in the Gunn-$r$ band, with a similar value for the bulge component. We adopt this mass-to-light ratio for both the bulge and disk components of the model galaxies. Next, this intrinsic luminosity must be converted to observed flux density, using the luminosity distance and $K$-correction. For a fixed observing bandpass and as-suming a spectral energy distribution for the galaxy which goes as $f_\nu \propto \nu^{-\alpha}$, the $K$-correction can be expressed as $-2.5(1 - \alpha)\log(1 + z)$. Using a typical $\alpha \sim 2$ (see Koo & Kron 1992), at a redshift of $z = 0.4$ this $K$-correction amounts to a dimming of 0.4 magnitudes, while at $z = 1$ the dimming is 0.75 magnitudes.

To simulate the instrumental response of WFC2 to the calculated incident flux from the galaxies, zeropoints from the WFC2 instrument manual (Burrows 1994) are used, assuming a $10^4$ second exposure in the F785LP filter. The raw image is then convolved with a WFC2 pointspread function generated by Tinytim Version 3 (Krist 1994). Finally, noise is added using the MKNOISE package in IRAF[1]. We assume a sky brightness of 22.2 mag arcsec$^{-2}$, representative of the ecliptic pole. To each pixel we add this sky background and a dark count background using a dark count rate of 0.016 e$^-$/pixel/sec and gain of 7 e$^-$/DN for the WFC2 CCDs. We add noise corresponding to Poisson noise on the background plus source counts, as well as readout noise corresponding to $5\sqrt{5}$ e$^-$ RMS, where we have assumed the total $10^4$ second exposure is broken into typically five shorter exposures. No artifacts due to cosmetic flaws on the CCD, hot pixels, flat-fielding errors, cosmic rays, or imperfect image registration are added.

3. **Results**

Figure 1 shows a set of synthesized images of the merging system at a redshift of $z = 0.4$. Note that this sequence is *not* the same as starting a merger at $z = 0.4$ and watching it evolve; in such a sequence the final time would correspond to a redshift of $z \sim 0.3$. Instead, Figure 1 shows how mergers at differing evolutionary phases would look at $z = 0.4$. Similarly, Figure 2 shows the same evolutionary phases observed at a redshift of $z = 1.0$. The sequence of images covers roughly 1.7 Gyr of time, beginning 150 Myr before the galaxies first collide (at $t = 24$) and continuing for 1 Gyr after the galaxies have merged (at $t = 65$).

An examination of Figure 1 reveals the rapid surface brightness evolution of the tidal debris associated with the merger. The tidal tails which form shortly after the first passage quickly expand and dis-perse, falling well below the sky brightness level *before* the galaxies actually merge. The total "visible" lifetime for the tails at this redshift is relatively short, $\sim 500$ Myr. After the galaxies merge, the remnant rapidly relaxes, as evidenced by the smooth isophotal contours in the main body in later images. The faint tidal features surrounding the remnant remain as the only morphological signature of the merging process. As this material wraps around the remnant, it blends smoothly in with the light profile of the remnant body, making it very hard to detect at late times. The remnant appears morphologically indistinguishable from a "typical" elliptical $\lesssim 1$ Gyr after the galaxies merge.

---

[1] IRAF is distributed by Kitt Peak National Observatory, National Optical Astronomy Observatories, operated by the Association of Universities for Research in Astronomy for the National Science Foundation.

At a redshift of $z = 1.0$ (Figure 2), the tidal features are even more difficult to detect. While the tails are still visible early in the interaction, they fade from view well before the actual merger. At this redshift, the tails are visible for less than 300 Myr. The loops and shells surrounding the remnant at late time are difficult to detect in these images, leaving no visible signatures of the merger only $\sim 500$ Myr after the galaxies merge. The lack of a clear merger signature at these higher redshifts suggests that the bulk of the morphological classification of these galaxies can only be done through the presence of companions, resulting in significant contamination in galaxy samples due to the presence of projected – rather than true – companions. Establishing a firm link between galaxy interactions and galaxy activity at high redshift, therefore, will be subject to considerable uncertainty.

Besides the visual appearance of the galaxy, quantitative structural parameters, such as brightness distributions which deviate from a $r^{\frac{1}{4}}$ law, variations in ellipticity, and the presence of isophotal twists, are often suggested as merger signatures. To quantify the image structure of the remnants in our images, we use the STSDAS ELLIPSE package to fit elliptical isophotes out to $\sim 2$ mag arcsec$^{-2}$ below sky for each of the $z = 0.4$ images (Figure 3). The efficiency of violent relaxation is evident; only 200 Myr after the merger, the surface brightness profile shows a good match to a $r^{\frac{1}{4}}$ profile. However, in the outer regions ($r > 0\rlap{.}''8$), isophotal twists and varying ellipticity can be seen. At these outer isophotes, the light is predominantly coming from the faint, irregular tidal debris, rather than the smooth distribution of galaxy light. As this tidal material falls back and wraps around the remnant, the isophotal anomalies fade with time, although some irregularities at large radius remain. However, when compared to "typical" elliptical galaxy profiles, the structural properties of the remnant show little evidence for the merging process only $\sim 500$ Myr after the galaxies merge.

Finally, for use in comparing with extant WFC1 data, we have created a third set of images (Figure 4), convolving the raw images with a typical WFC1 PSF and using the instrumental characteristics of the WFC1 CCDs (MacKenty 1992). The much poorer image quality of these images renders the faint structure of the tidal features nearly undetectable. The long tidal tails disappear shortly after their formation, and the isophotal contours in the remnant are quite smooth almost immediately upon merging. While deconvolution can restore some resolution in the image centers, the faint tidal debris – the only merger signature remaining – has been lost. Therefore, WFC1 data will not be able to identify signatures of a galaxy merger almost immediately after the galaxies have merged.

## 4. Discussion

Studies which seek to connect the occurrence of mergers and interactions with the Butcher-Oemler effect and formation of E+A galaxies are necessarily subject to selection effects and incompleteness due to the morphological criteria employed and the longevity of these signatures. The modeled images presented here indicate that the lifetime over which HST WFC2 imaging may identify a galaxy as a merger remnant is $\lesssim 1$ Gyr at $z = 0.4$, and much shorter, $\lesssim 500$ Myr, at $z = 1.0$. With WFC1 data, the identification of merger remnants is nearly impossible shortly after merging, as the low surface brightness tidal features are lost in the aberrated PSF and not recovered by deconvolution.

The models analyzed here involve only stellar-dynamical evolution of merging disk galaxies; the effects of gas dissipation and star formation are not included. The intense starbursts often associated with galaxy mergers occur predominantly in the central regions of the merging galaxies, and die out shortly after the galaxies merge (e.g., Mihos, Richstone, & Bothun 1992; Mihos & Hernquist 1994a). Star formation in the tidal debris is generally suppressed by the interaction after the initial collision, save for in a few small star-forming clumps. The tidal features will thus redden and fade *intrinsically* with time, as their stellar populations evolve. This fading can amount to $\sim 1$ $R$ magnitude 1 Gyr after the truncation of star formation; accordingly, these features may actually be much fainter than indicated in the simulated images, further complicating their detection. However, the central starburst population formed during the merger may be detectable via deviations from a pure $r^{\frac{1}{4}}$ law in the central regions of the remnant (Mihos & Hernquist 1994b,c).

The visibility of tidal tails could be enhanced through favorable combinations of orbital geometry and viewing angle. When the orbital and spin angular momentum vectors of the galaxies are aligned, the tails will be predominantly two-dimensional (Toomre & Toomre 1972). Such tails viewed edge-on will have a higher surface brightness than those seen in the simulated images. Thus the brightest examples of long tidal tails should be linear in appearance, similar to that shown by the nearby interacting system NGC 4676 (e.g., Arp 1966). More generally, however, tidal tails will have more complicated three-dimensional structure, and when viewed along random lines of sight they should appear much like the tails shown in these synthesized images.

Several recent HST investigations have examined the morphology of starbursting and peculiar galaxies in clusters at intermediate redshift. Dressler et al. (1994a,b) identified a significant fraction of starburst galaxies in the $z = 0.41$ cluster CL 0939+4713 as merging objects, based on distorted or amorphous appearance. Because of the rapid relaxation associated with the merging, such a classification scheme will select objects which are dynamically very young, within a few hundred Myr of the merger event. It is at these times that starburst activity in merging galaxies is highest (e.g., Mihos et al. 1992, Mihos & Hernquist 1994a). Because the starburst phase corresponds closely to the morphologically peculiar phase, studies which aim to correlate the two (e.g., Forbes et al. 1994) should suffer from fewer incompleteness effects, although classification selection effects and contamination due to projection will still be important.

Looking for the effects of mergers over longer timescales, however, is more difficult due to the rapid relaxation of the remnant, and the faintness of the tidal debris. Using WFC1, Couch et al. (1994) found that the E+A systems in clusters AC 114 ($z = 0.31$) and Abell 370 ($z = 0.37$) showed no visible signs of a merger. However, the lifetime of this post starburst E+A phase ($\sim$ 1–2 Gyr, Couch et al.) is much longer than the observable lifetime of the merger signatures as viewed by WFC1. In fact, even if *all* E+A galaxies were spawned through galaxy mergers, the brief lifetime for the tidal debris suggests that only a very small fraction of those E+A systems would show merger signatures. Therefore, the fact that the E+A systems in AC 114 and Abell 370 appeared morphologically normal in WFC1 images is not inconsistent with a merger origin for those galaxies.

A more quantitative approach to measuring image structure in E+A galaxies was employed by Wirth et al. (1994) in WFC1 images of the $z = 0.55$ cluster 0016+16. Defining a concentration index $C = 5\log(r_{0.8}/r_{0.2})$ for the light profile, Wirth et al. found that the concentration indices for the E+A galaxies in 0016+16 were more indicative of disk systems rather than $r^{\frac{1}{4}}$-law systems. Because the outermost isophotes in the merger models drop off slightly more quickly than in a pure $r^{\frac{1}{4}}$-law system, it is possible that the concentration index $C$ for the mergers could be reduced to values more typical of disk systems. To test this possibility, G. Wirth kindly performed a independent measure of $C$ for the merger remnant images. In the images shown in Figure 1, and in additional images mimicking WFC1 observations at $z = 0.55$, the $C$ values for the merger images were indistinguishable from a pure $r^{\frac{1}{4}}$-law model, while an exponential disk model and the E+A galaxies in 0016+16 showed significantly lower values of $C$. The "disky" E+A galaxies in 0016+16, therefore, likely represent a population of objects distinct from those produced by mergers of disk galaxies.

With the improved optics of WFC2, however, a more stringent test of the merger origin for E+A galaxies may be possible. Under ideal conditions, the low surface brightness debris left over from the merging process should be visible over timescales more comparable to the lifetime of the E+A phase. Because of their faintness, however, relatively long exposures, low sky background, and good flat-fielding will be necessary to reveal these features. As such, even a modest fractional detection rate would be a very strong indication that disk galaxy mergers are driving the formation of E+A galaxies in intermediate redshift clusters.

I thank Lars Hernquist for making available the numerical models analyzed in this work, and Greg Wirth for calculating the concentration indices for the synthetic images. I also thank Drew Phillips,

Matt Bershady, and Duncan Forbes for many helpful discussions. This work was supported in part by the Pittsburgh Supercomputing Center, NASA Theory Grant NAGW–2422, and the NSF under Grants AST 90–18526 and ASC 93-18185.

Fig. 1.— a) Synthesized WFC2 images at a redshift of $z = 0.4$. Time is shown in the upper left corner, with unit time corresponding to 13 Myr. Each subframe measures $15''$ on a side. The images are scaled to best bring out faint tidal features. b) Contour map. Contours begin at $1\sigma$ above sky and are spaced at 1.0 magnitude per square arcsec intervals.

Fig. 2.— a) Synthesized WFC2 images at a redshift of $z = 1.0$. Each subframe measures $5''$ on a side. b) Contour map.

Fig. 3.— Derived image structure properties as a function of radius for the $z = 0.4$ WFC2 images. a) surface brightness profile, b) ellipticity, c) major axis position angle.

Fig. 4.— a) Synthesized WFC1 images at a redshift of $z = 0.4$. Each subframe measures $15''$ on a side. b) Contour map.